\documentclass[prd,a4paper,aps,twocolumn,
               nofootinbib,nobibnotes,superscriptaddress,preprintnumbers]
              {revtex4}

\usepackage[dvips]{graphicx}
\usepackage{amsmath,amssymb,mathrsfs}
\usepackage{bm}
\usepackage{times}
\usepackage{epsfig}
\usepackage{verbatim}
\usepackage{bm}

\usepackage{graphics}
\usepackage{graphicx,epsfig,amssymb,amsmath,color, cancel}

\newcommand{\be}{\begin{equation}}
\newcommand{\ee}{\end{equation}}
\newcommand{\ba}{\begin{array}}
\newcommand{\ea}{\end{array}}
\newcommand{\bea}{\begin{eqnarray}}
\newcommand{\eea}{\end{eqnarray}}

\newcommand{\cms}{{\rm cm}^3/{\rm s}}

\begin{document}

\preprint{CERN-PH-TH/2013-068, MCTP-13-10}

\title{Higgsogenesis}

\author{G\'eraldine Servant}
\affiliation{CERN Physics Department, Theory Division, CH-1211 
Geneva 23, Switzerland,}
\affiliation{ICREA at IFAE, Universitat Aut\`onoma de Barcelona, 08193 Bellaterra, Barcelona, Spain,}
\affiliation{Institut de Physique Th\'eorique, CEA/Saclay, F-91191 Gif-sur-Yvette C\'edex, France}
\author{Sean Tulin}
\affiliation{Department of Physics, University of Michigan, Ann Arbor, MI 48109  }

\date{\today}

\begin{abstract}

In addition to explaining the masses of elementary particles, the Higgs boson may have far-reaching implications for the generation of the matter content in the Universe. For instance, the Higgs plays a key role in two main theories of baryogenesis, namely electroweak baryogenesis and leptogenesis.  In this letter, we propose a new cosmological scenario where the Higgs chemical potential mediates asymmetries between visible and dark matter sectors, either generating a baryon asymmetry from a dark matter asymmetry or vice-versa.
We illustrate this mechanism with a simple model with two new fermions coupled to the Higgs and discuss associated signatures.

\end{abstract}

\maketitle

{\bf Introduction:}  Asymmetries between particles and antiparticles in the thermal plasma of the early Universe play an important role in cosmology, with the most notable example being the baryon asymmetry that we observe today.  If the baryon asymmetry arose before the electroweak phase transition (EWPT), then necessarily other asymmetries must have existed as well due to chemical equilibrium~\cite{Harvey:1990qw,Chung:2008gv}.  Electroweak sphalerons convert asymmetries between baryon and lepton number, while Yukawa interactions can induce a Higgs asymmetry between $H$ and $H^\dagger$ (since the Higgs doublet $H$ is a complex scalar).  A lepton asymmetry plays a key role in leptogenesis scenarios, where lepton number is generated dynamically and gets converted into baryon number~\cite{Davidson:2008bu}.  Leptogenesis may also source an asymmetry within a hidden sector that ultimately provides the dark matter (DM) density of the Universe within the framework of asymmetric DM~\cite{Falkowski:2011xh}.  

In light of the recent Higgs discovery, it is tempting to ask under which circumstances the {\it Higgs asymmetry} could prevail to generate the relic abundance of baryons or DM.  We dub this generic scenario {\it Higgsogenesis}.  An important difference from leptogenesis is that Higgs charge is rapidly erased after the EWPT --- since the Higgs vacuum expectation value (vev) violates Higgs number --- as opposed to lepton number, which is frozen in.  Asymmetries must be decoupled from the Higgs at this point or else they may be erased. 

Although Higgsogenesis may stand alone as a baryogenesis mechanism, in this letter we pay special attention to the possible connection to DM.  If DM is associated with new physics at the electroweak symmetry breaking scale, one typically expects couplings between the Higgs and DM.  A Higgs asymmetry can be transferred to the dark sector and would naturally imply that DM is asymmetric (if it is not self-conjugate). This case was proposed for Higgsino DM in a finely-tuned supersymmetric framework~\cite{Blum:2012nf}.   Inversely, a primordial asymmetry produced in the dark sector could be transferred to the visible sector, therefore leading to a theory of baryogenesis that does not require violation of baryon or lepton number beyond the Standard Model (SM), relying on sphalerons only.


{\bf Model:} To show how the Higgs can play a role in generating matter asymmetries, we consider a simple, illustrative model.  We introduce an EW singlet $X_1$ and a doublet \mbox{$X_2 \equiv (X_2^+,X_2^0)$}, both vector-like Dirac fermions and uncharged under color.  DM is (mostly) $X_1$, while $X_2$ is needed to transfer asymmetries between the visible sector and the ``$X$ sector.''  We also introduce a gauge-singlet dark mediator $\phi$, which can be scalar or vector and which is required for DM annihilation.  

The Lagrangian includes vector masses $m_{1,2}$ for $X_{1,2}$ and a Yukawa coupling with the SM Higgs doublet $H$:
\be
- \mathscr{L}  \supset m_1 \bar{X}_1 X_1 + m_2 \bar{X}_2 X_2  +  y_H \left(\bar{X}_2 X_1 H + \textrm{h.c.}  \right) \, . \label{yukawa}
\ee
We suppose that $X_{1,2}$ respect an $X$-number symmetry, which forbids a Majorana mass for $X_1$.  
We also consider nonrenormalizable dimension-five operators that violate $X$-number:
\be
- \mathscr{L} \supset \frac{1}{\Lambda_1} |H|^2 X_1^2 +  \frac{1}{\Lambda_2} (H^\dagger X_2)^2 + \textrm{h.c.} \label{dim5}
\ee
We refer to the $\Lambda_2$ term as the {\it Higgs transfer operator} since it equilibrates charge between the Higgs and $X$ sector; the $\Lambda_1$ term simply leads to $X$-number washout.  The $X$-number symmetry prevents mixing of $X_{1,2}$ with neutrinos and guarantees DM stability.   However, since the dimension-5 operators violate it, it would have to be broken softly.  (Alternately mixing with neutrinos can be forbidden by imposing $U(1)_{B-L}$ if it is broken by 2 units.) 

After electroweak symmetry breaking, the Higgs vev $v \equiv \langle H^0 \rangle$ in Eqs.~\eqref{yukawa} and \eqref{dim5} causes the two Dirac neutral gauge eigenstates \mbox{${X}_{1}$, ${X}_2^0$} to mix and split, respectively, into four Majorana mass eigenstates.  As we discuss below, our mechanism is viable only for small mixing angle $\theta = \tfrac{1}{2} \tan^{-1}[  2y_H v/(m_2 \!-\! m_1)] \ll 1$.  In this limit, the mass eigenvalues are approximately $m_{1} \pm \delta_{1}$ and $m_{2} \pm \delta_{2}$, where $\delta_{1,2} \approx v^2/\Lambda_{1,2} +  v^2 \theta^2 /\Lambda_{2,1}$.
The Majorana splittings $\delta_{1,2}$ lead to particle-antiparticle oscillations that can erase an $X$ asymmetry after the EWPT
\cite{Cirelli:2011ac,Tulin:2012re,Buckley:2011ye}.\footnote{We (ab)use the notation $X_{1,2}$ to denote {\it both} the gauge and mass eigenstates, since they are approximately equal for $\theta\ll 1$ and $\delta_{1,2} \ll m_{1,2}$.}


{\bf Origin of matter:}  Higgs charge provides a portal through which asymmetries in the visible and $X$ sectors can be linked.  There are two generic cases, depending on whether an asymmetry flows from the visible sector to the $X$ sector, or vice versa.  We consider both cases below, but first we discuss some preliminaries.

Equilibration of charge asymmetries 
between different species is governed by relations between chemical potentials, assuming various interactions are in chemical equilibrium~\cite{Harvey:1990qw,Chung:2008gv}.  
For species $i$, the charge density is $n_i  \equiv n^+_{i} \!-\! n^-_{i} = T^2 k_i(m_i/T) \mu_i/6$, where $T$ is temperature, $\mu_i$ is the chemical potential (with $\mu_i \ll T$), and 
$k_i(x) \equiv 6 \pi^{-2}\,  g_i  \int_x^\infty dz \, z \sqrt{z^2 - x^2} \, e^z/(e^z \pm 1)^2$,
with $+$ ($-$) for fermions (bosons), and $g_i$ counts color, generation, etc.  (Note $k_i(0) = g_i$ for chiral fermions, $k_i(0) = 2g_i$ for complex scalars.)  For $T \lesssim 10^6$ GeV, all SM interactions are in chemical equilibrium~\cite{Nardi:2005hs}, giving Yukawa and weak sphaleron equilibrium conditions (before the EWPT)
\begin{align} \label{yukawachem}
\frac{n_Q}{k_Q} + \frac{n_H}{k_H} &= \frac{n_u}{k_u}  \, , & \frac{n_Q}{k_Q} - \frac{n_H}{k_H} &= \frac{n_d}{k_d}  \, , \\
\nonumber
\frac{n_L}{k_L} - \frac{n_H}{k_H} &= \frac{n_e}{k_e} \, , & \frac{3n_Q}{k_Q} + \frac{n_L}{k_L} &= 0 \, , 
\end{align}
where $Q,L$ are the left-handed quark and lepton doublets, and $u,d,e$ are the right-handed quarks and charged leptons, summed over three generations.\footnote{Having $X_2$ charged under $SU(2)_L$ modifies the sphaleron equilibrium condition at high $T$, with chiral states $X_{2L}$ and $(X_{2}^C)_L$ participating in the sphaleron vertex.  Ref.~\cite{Ibanez:1992aj} estimated that for $T \ll \sqrt[3]{m_2^{2} M_{\rm Pl}}$, their effect in the sphaleron may be neglected due to washout from the $X_2$ mass term.}   Approximating SM states to be massless, the $k$-factors are
\be
k_Q = 18 , \; k_u = 9, \; k_d = 9, \; k_L = 6, \; k_e = 3, \; k_H = 4 \; .
\ee
We also set the hypercharge density to zero:
\be \label{hypercharge}
Y = \frac{n_Q}{6} + \frac{2n_u}{3}  - \frac{n_d}{3} - \frac{n_L}{2}  - n_e + \frac{n_H}{2} + \frac{n_{X_2}}{2} = 0  \; .
\ee
Lastly, we consider chemical equilibrium conditions from $H$ couplings to the $X$ sector.  The $y_H$ term in Eq.~\eqref{yukawa} gives
\be \label{eq:npyuk}
\frac{n_{X_1}}{k_{X_1}} + \frac{n_H}{k_H} = \frac{n_{X_2}}{k_{X_2}} \; ,
\ee
provided $X_2 \leftrightarrow X_1 H$ is faster than the Hubble rate.
If the nonrenormalizable terms in Eq.~\eqref{dim5} are in equilibrium, we have
\be
\frac{n_{X_1}}{k_{X_1}} = 0 \, , \qquad \frac{n_{X_2}}{k_{X_2}} = \frac{n_{H}}{k_{H}} \, . \label{dim5chem}
\ee
Since the scattering rate for, e.g., $H X_2 \to H^{\dagger} \bar X_2$ scales as $T^3/\Lambda_2^2$, these interactions are in equilibrium for $T \gtrsim \Lambda_{1,2}^2 /M_{\rm Pl}$.  For lower temperatures (or if $\Lambda_{1,2}$ are sufficiently large), Eqs.~\eqref{dim5chem} do not hold --- an important fact depending on which way the asymmetry is transfered.  
%
\begin{figure*}[t!]
\begin{center}
\includegraphics[scale=0.74]{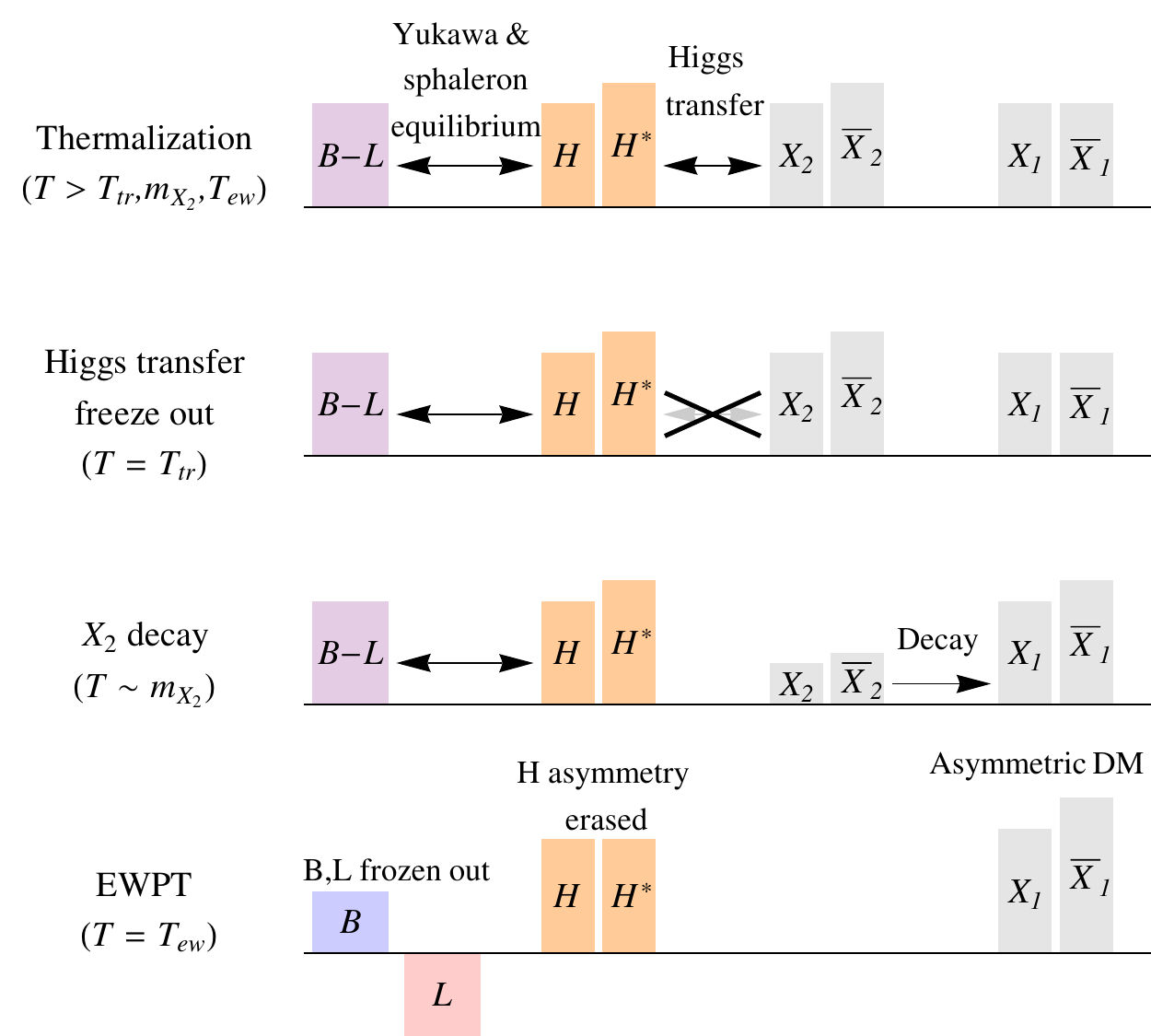}
\hspace{.9cm}
\includegraphics[scale=0.62]{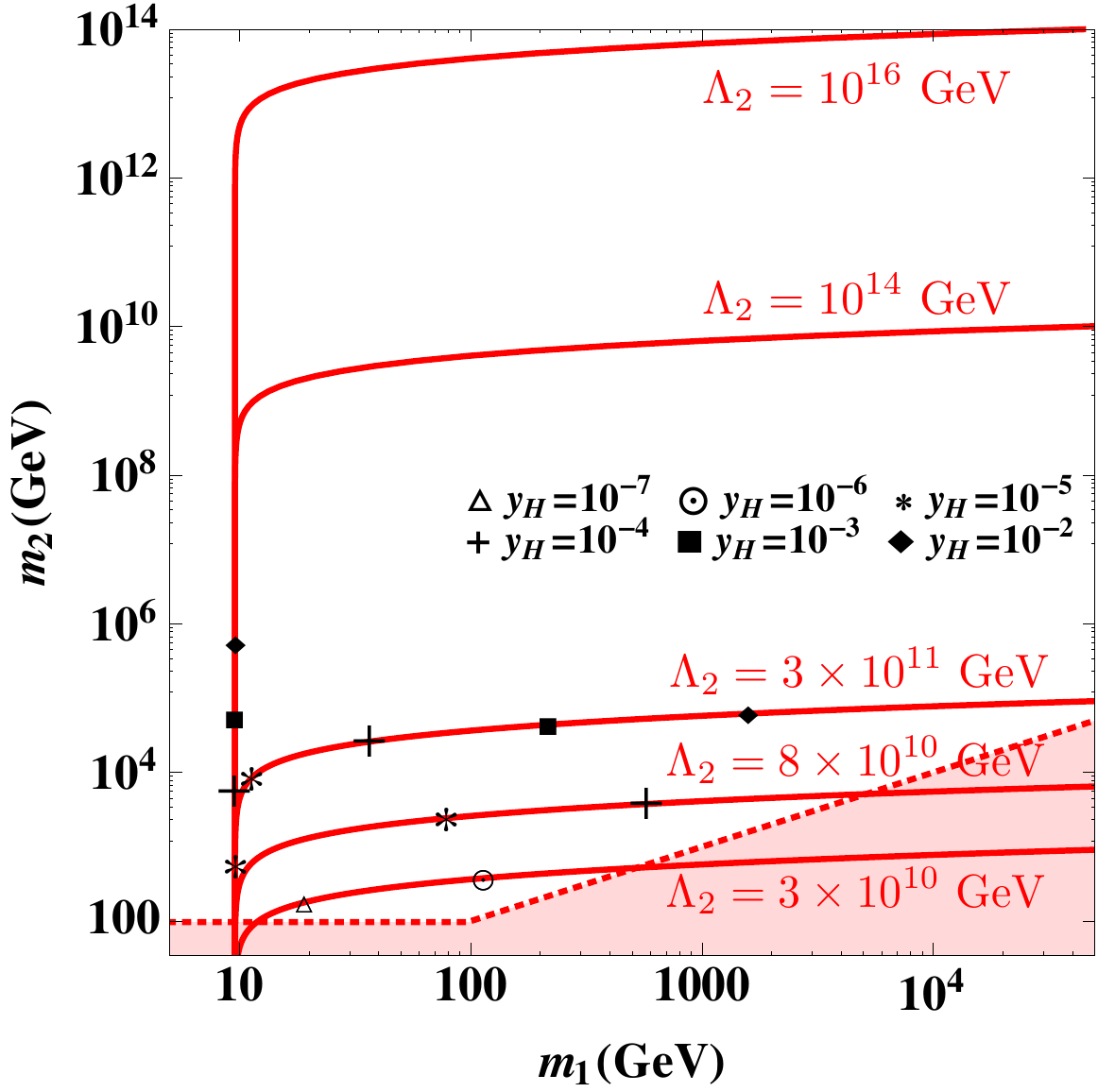}
\caption{ \small Left: Schematic representation of the charge transfer dynamics of case I.  A primordial $B\!-\!L$ charge generates a Higgs asymmetry, which subsequently flows to $X_2$ by the Higgs transfer operator.  When this operator freezes out at $T_{tr}$, the visible and $X$ sectors are no longer in chemical equilibrium and $X$ charge is frozen in.  Later, $X_2 \to X_1 H$ decays transfer the DM asymmetry to $X_1$. Right: For case I, the red solid lines represent the contours for the correct DM relic abundance for given $\Lambda_2$. The shaded area is excluded to guarantee $m_1<m_2$ and $m_2 \gtrsim 100$ GeV.
For the smallest $\Lambda_2$ values that lead to a large range of DM masses at the (sub)TeV scale and
$m_1 \lesssim x_{f} T_{ew}\sim 3$ TeV, the symbols on the contours indicate the $y_H$-dependent lower bounds on $(m_1,m_2)$ for $X_1$-$\bar{X}_1$ oscillations to start after $ X_1$ freeze-out. \label{fig:diagcase1} }
\end{center}
\end{figure*}
%

{\it Case I --- Asymmetry from the visible sector:}  First, we consider an asymmetry that is transfered to the $X$ sector from the visible sector.  We suppose a primordial $B\!-\!L$ asymmetry is generated before the EWPT, via leptogenesis~\cite{Davidson:2008bu} or another mechanism, inducing a nonzero chemical potential for $H$.  Higgs charge flows to the $X$ sector through the $(H^\dagger X_2)^2$ operator, assumed to be in chemical equilibrium.  
Solving Eqs.~(\ref{yukawachem}-\ref{dim5chem}), the $X_{1,2}$ charge densities are
\be
n_{X_1} = 0 \, , \quad n_{X_2} = - \frac{16 k_{X_2}}{13 k_{X_2} + 316} \, n_{B-L} \, ,
\label{eq:case1master}
\ee
where $n_{B - L}$ is the primordial $B\!-\!L$ density.  (These relations hold independently of whether or not the $|H|^2 X_1^2$ and $y_H \bar{X}_2 X_1 H$ operators are in equilibrium, since there is no source term for $X_1$ charge.)  Eventually, Higgs charge transfer freezes out, at temperature $T_{tr} \sim \Lambda_{2}^2/M_{\rm Pl}$, freezing in the $X$ asymmetry.  Below $T_{tr}$, $X$ charge gets converted from $X_2$ to $X_1$ as $X_2$ becomes nonrelativistic and decays $X_2 \to X_1 H$.  (We assume the $|H|^2 X_1^2$ operator is also frozen out at this point, otherwise $n_{X_1}$ washed out.)  This dynamics is summarized in Fig.~\ref{fig:diagcase1}

Next, at the EWPT (at $T_{ew} \sim 100$ GeV), the baryon asymmetry freezes out, giving $n_B = (28/79) n_{B-L}$. (This relation assumes a first-order EWPT; our predictions will be modified by $\mathcal{O}(1)$ if this is not the case.)  Thus, the DM-to-baryon charge ratio is
\begin{align}
\frac{n_{X_1}}{n_B} = - \frac{316\, k_{X_2}^{ tr}}{2212+ 91 k_{X_2}^{ tr}} 
\approx \left\{ \begin{array}{cc} -0.5 & x_{ tr} \lesssim 1 \\
 -0.4 \, x_{ tr}^{3/2}\, e^{-x_{\rm tr}} & x_{ tr} \gg 1 \end{array} \right.
\end{align}
where $x_{tr} \equiv m_2/T_{tr}$ and $k_{X_2}^{ tr} \equiv k_{X_2}(x_{tr})$.  
Eventually, the symmetric $X_1, \bar X_1$ density annihilates efficiently (discussed below), leaving behind the residual asymmetric component as DM.  The ratio of DM-to-baryon energy densities observed today is ${\Omega_{dm}}/{\Omega_{b}} = (m_{1}/m_p) \,  \left|n_{X_1}/n_B \right| \approx 5$.  The maximal $X$ charge arises if $X_2$ is relativistic at $T_{tr}$ ($x_{tr} \lesssim 1$), giving a lower bound for the DM mass, $m_1 \approx 10$ GeV.  A smaller $X$ charge occurs for $x_{tr} \gg 1$, requiring larger $m_1$.
Fig.~\ref{fig:diagcase1} shows contours of $(m_1,m_2)$ giving the correct relic abundance for different $\Lambda_2$.

A nontrivial complication is the fact that Eq.~\eqref{dim5} generates Majorana masses $\delta_{1,2}$ for $X_{1,2}$ that turn on at the EWPT.  These terms violate $X$-number and lead to oscillations that erase the $X$ asymmetry.  To realize asymmetric DM, we require (i) $X_2\leftrightarrow\bar X_2$ oscillations do not occur before $X_2$ has transfered its charge to $X_1$, and (ii) $X_1 \leftrightarrow \bar X_1$ oscillations do not occur before freeze-out of $X_1 \bar X_1$ annihilation.   Condition (i) is generally satisfied since $X_2$ undergoes rapid gauge scatterings with SM fermions $f$ in the plasma, which delays oscillations~\cite{Cirelli:2011ac,Tulin:2012re}.  The gauge scattering rate \mbox{$\Gamma_s( X_2 f \to X_2 f) \sim  G_F^2 T^5$} is larger than the oscillation rate $\delta_2$ for $T \!\gtrsim \!{\rm GeV} \!\times\! (10^{11} \, {\rm GeV}/\Lambda_2)^{1/5}$, assuming $\Lambda_2 \ll \Lambda_1/\sin^2\theta$.  Since $m_2 \gtrsim 100$ GeV is required by LEP2 chargino searches~\cite{Beringer:1900zz}, $X_2$ has decoupled from the plasma when its oscillations begin.  Condition (ii) is less easily satisfied.  $X_1 \leftrightarrow \bar X_1$ oscillations turn on at temperature $T_{osc} \sim {\rm min}(T_{ew}, \, \sqrt{ \delta_1 M_{pl} })$, with no gauge scattering to delay their onset.  The ``min'' corresponds to the fact that $\delta_1$ is proportional to the Higgs vev and becomes nonzero only after the EWPT.  If $\Lambda_{1} \sim \Lambda_2$ and $T_{tr} < M_{pl}$, then $\delta_1 \gtrsim v^2/M_{pl}$ and $T_{osc} \sim T_{ew}$.  The only way to avoid erasure of the $X_1$ asymmetry is if annihilation freezes out before $T_{ew}$, requiring either multi-TeV DM or an unusually small $T_{ew} \ll 100$ GeV~\cite{Blum:2012nf}.  On the other hand, if $\Lambda_1 \gg \Lambda_2$, then $\delta_1$ is suppressed by $\theta \ll 1$.
In this case, imposing that $X_1$ freezes out before oscillating leads to lower bounds on $m_1$ and $m_2$, which depend on the value of the Yukawa coupling $y_H$. This is illustrated in Fig.~\ref{fig:diagcase1}, where the symbols on the red contours indicate the lowest allowed masses (for a given $y_H$) for the asymmetry not to be erased by oscillations. There is no such bound for $m_1 \gtrsim x_{f} T_{ew}\sim 3$ TeV.  

\begin{figure*}[t!]
\begin{center}
\includegraphics[scale=0.78]{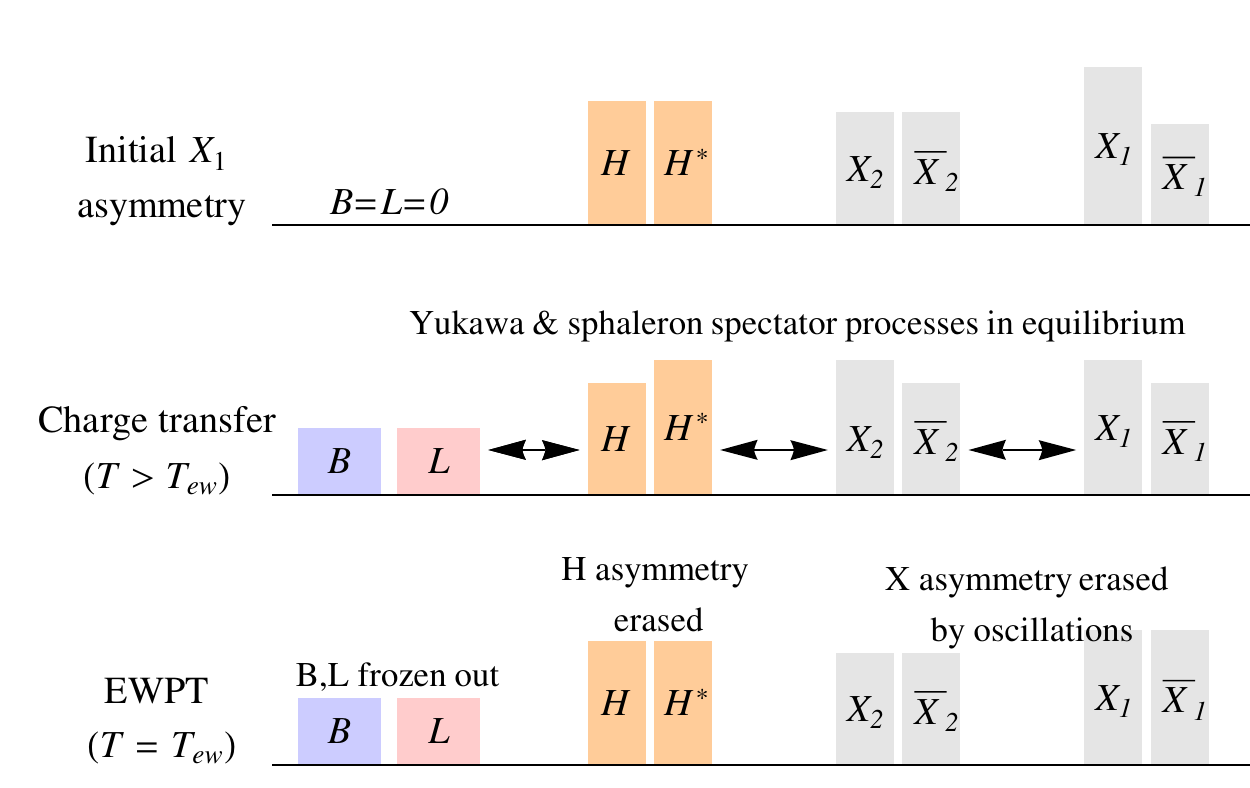}
\hspace{.9cm}
\includegraphics[scale=0.58]{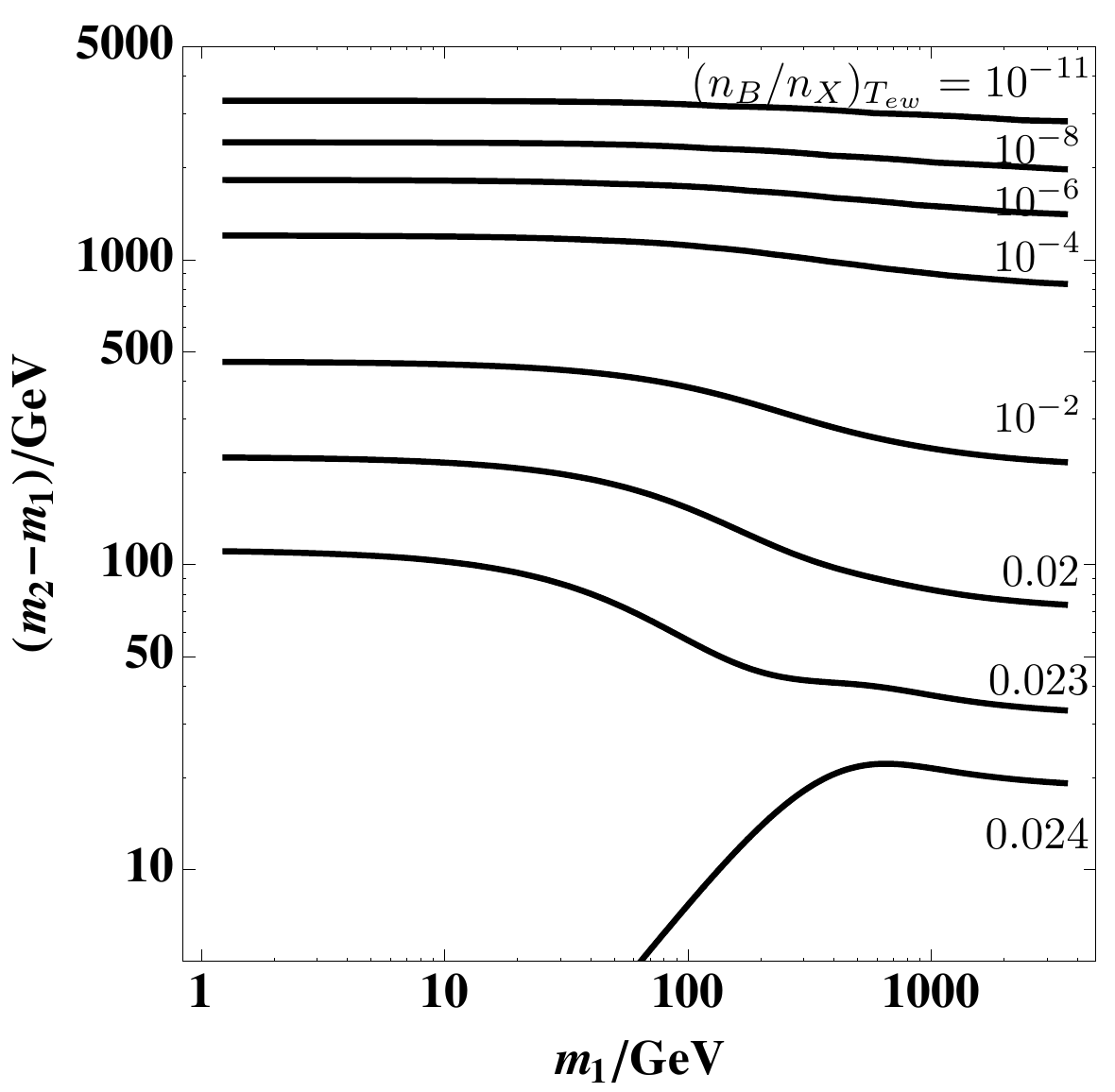}
\caption{ \small Left: Schematic representation of the charge transfer dynamics of case II.  A primordial $X_1$ charge generates a Higgs asymmetry, which through spectator processes, biases electroweak sphalerons into generating $B$ and $L$ charge (even though $B\!-\!L=0$).  The $B$ density is frozen in for a sufficiently strong first order EWPT.  Particle-antiparticle oscillations washout the $X$ asymmetry after the EWPT, and DM is symmetric at thermal freeze-out.
Right: For case II,  contours of $\left(n_B/{n_X}\right)_{T_{ew}}$ at the EWPT, as a function of $m_{1,2}$.  The final baryon asymmetry today is $\left(n_B/{n_X}\right)_{T_{ew}}$ times a washout factor $W$ (see text).  \label{fig:diagcase2} }
\end{center}
\end{figure*}

{\it  Case II --- Asymmetry from the $X$ sector:}  Next, we consider an alternative case where an asymmetry from the $X$ sector is transferred to the visible sector, thereby generating the $B$ asymmetry.  At some initial time an $X$ asymmetry is generated (e.g., a heavy scalar may decay out-of-equilibrium, with CP-violating rates for $X_1 X_1, \,  \bar{X}_1 \bar{X}_1$ final states).  The $X_1$ asymmetry generates a chemical potential for $H$, which flows to the visible sector through Yukawa and sphaleron interactions (see Fig.~\ref{fig:diagcase2}).  As before, the resulting $B$ asymmetry is determined by requiring these interactions to be in chemical equilibrium (with $Y=0$), given by Eqs.~(\ref{yukawachem}-\ref{eq:npyuk}).  We have
\be
\frac{n_B}{n_X} = \frac{12 k_{X_2}}{13 k_{X_1} k_{X_2} + 316 (k_{X_1} + k_{X_2})} \; ,\label{nBratio2}
\ee
even though $B\!-\!L$ is zero.  
We also require that the dimension-five operators are {\it not} in equilibrium, which otherwise would wash out this asymmetry. That is, we do not impose Eqs.~\eqref{dim5chem}; otherwise the only solution is $n_X = n_B = 0$.

The $B$ asymmetry freezes-out at the EWPT.  In the limit that the EWPT is instantaneous, the $B$-to-$X$ charge ratio is fixed by Eq.~\eqref{nBratio2} at $T_{ew}$, given by
\be
\left( \frac{n_B}{n_X}\right)_{T_{ew}} \approx \left\{ \begin{array}{cc} 
0.024 & m_{1,2} \ll T_{ew}\\
0.076 \left(\frac{m_2}{m_1}\right)^{3/2} e^{- \frac{m_2-m_1}{T_{ew}}}  & m_{1,2} \gg T_{ew} \end{array} \right. .
\ee
Values of $n_B/n_X$ at ${T_{ew}}$ are shown in Fig.~\ref{fig:diagcase2}.

The finite duration of the EWPT causes additional washout of $n_B$.  Since $\mu_H$ is rapidly relaxed to zero during the EWPT (since the vacuum violates Higgs number), the $B$ asymmetry also relaxes away if sphalerons are still active.  The washout factor $W$ has been calculated from the finite temperature sphaleron rate after the EWPT to be
$W \approx \exp( - 10^{10} \kappa \zeta^7 e^{-\zeta} )$
where $\kappa \sim 0.001$ is the fluctuation determinant (for $m_h = 125$ GeV)
and $\zeta = E_{\rm sph}(T_c)/T_c$ gives the sphaleron barrier energy at the critical temperature $T_c$~\cite{Quiros:1999jp}.
%
%
%
%
%

Ultimately, the baryon asymmetry today is 
\be
{n_B}/{s} = W \left( {n_B}/{n_X} \right)_{T_{ew}} \left({n_X}/{s} \right)_{in} \; , \label{nBtoday}
\ee
where $\left({n_X}/{s} \right)_{in}$ is the initial $X$ charge asymmetry, and $s$ is the entropy density.  Since $({n_B}/{n_X})_{T_{ew}} \lesssim 10^{-2}$ and $\left({n_X}/{s} \right)_{in} \lesssim g_*^{-1} \sim 10^{-2}$, we require $W \gtrsim 10^{-6}$ to achieve $n_B/s \approx 10^{-10}$.  This leads to a similar condition as in EW baryogenesis, $v(T_c)/T_c  \gtrsim 1$, which is only weakly sensitive to $W$.  The dark mediator $\phi$, introduced for annihilation, can in principle play a role to strengthen the EWPT as well.

Since $n_B$ is much smaller that the asymmetry in the $X$ sector, oscillations are crucial for erasing the latter and obtaining the correct $\Omega_{ dm}$.  Oscillations begin at $T_{osc} \sim T_{ew}$ for $\Lambda_1 \lesssim M_{\rm Pl}$, the DM asymmetry is erased before freeze-out, and $\Omega_{ dm}$ is determined by symmetric freeze-out by requiring $\langle \sigma v \rangle \approx  6 \times 10^{-26} \, {\rm cm}^3/{\rm s}$.
At the same time, we require that the initial $X$ asymmetry is generated at $T \ll \Lambda_{1,2}^2/M_{\rm Pl}$, such that dimension-five interactions are never in equilibrium. We find that the condition of having enough oscillations -- without equilibrating the asymmetries away -- is satisfied for $\Lambda_{1,2} \gg 4 \times 10^{10}$ GeV and $m_1\ll 10^8$ GeV.

{\bf Symmetric annihilation \& phenomenology:} For asymmetric freeze-out (case I), $X_1 \bar X_1$ annihilation must be efficient enough to deplete the symmetric density, requiring $\langle \sigma v \rangle \gtrsim 6 \times 10^{-26} \, \cms$~\cite{Kaplan:2009ag}.  For symmetric freeze-out (case II), the lower limit must be saturated to give the correct relic density.  In principle, $X_1 \bar X_1$ can annihilate into SM states directly through gauge interactions for $\theta \sim 1$.  However, this also leads to a sizable spin-independent (SI) cross section for $X_1$ scattering with the neutron ($n$) through $Z$ exchange: 
$\sigma_{n}^{\rm SI} \approx {\mu_n^2 G_F^2 \sin^4 \theta}/(2\pi) \approx 7 \times 10^{-39} \, {\rm cm^2} \, \sin^4  \theta \, ,$
where $\mu_n \approx m_n$ is the reduced mass.  Current XENON100 limits require $\theta < 0.1$ for the range $10 < m_1 < 10^4$ GeV (this limit is a function of mass, with the strongest limit at $m_1 = 55$ GeV requiring $\theta < 0.03$)~\cite{Aprile:2012nq}.  For small values of $\theta$, achieving a large enough $\langle \sigma v \rangle$ is excluded.

The presence of a dark mediator $\phi$ provides a means of efficient annihilation through the $t$-channel process $X_1 \bar X_1 \to \phi\phi$ for $m_\phi < m_1$.  At leading order in the relative velocity $v$, the cross section is
$\sigma v \approx \pi \alpha_X^2 c(v)/m_1^2$, where $\alpha_X$ is the coupling, and $c = 1$ if $\phi$ is a vector or $c = 3v^2/8$ ($v^2/24$) if $\phi$ is a (pseudo)scalar.  A wide range of $(m_1, \alpha_X)$ can achieve a sufficient cross section, although a larger coupling is required for the scalar cases due to the $p$-wave suppression.  

Electroweak $X_2$ pair production can be studied at colliders, provided it is kinematically accessible.  The dominant decay modes are $X_2^+ \to W^{(*)} X_1$ and $X_2^0 \to Z^{(*)} X_1$, with $X_1$ escaping as missing transverse energy (MET).  Recent CMS and ATLAS analyses at 8 TeV (with 9 fb$^{-1}$ and 21 fb$^{-1}$, respectively) have searched for $3 \ell + {\rm MET}$ final states characteristic of $X^+_2 \bar X_2^0$ production~\cite{CMS-PAS-SUS-12-022,ATLAS-CONF-2013-035}, with ATLAS excluding $m_2 \lesssim 320$ GeV for $m_1 \lesssim 70$ GeV.  $X_2^0 \bar X_2^0 \to X_1 \bar{X}_1 Z^{(*)} Z^{(*)}$ can be studied in $4 \ell + {\rm MET}$ searches.


Due to particle-antiparticle oscillations, annihilation can occur today, producing an observable signal in DM halos, while annihilation at earlier times can modify reionization as imprinted on the cosmic microwave background~\cite{Lopez-Honorez:2013cua}.  The specific indirect and direct detection signals depend on the spin and CP of $\phi$, and how it couples to the SM~\cite{Jackson:2013pjq}, with additional possible correlations with electric dipole moment searches~\cite{Fan:2013qn}. 
Meditors with highly suppressed couplings to the SM can be still be probed through astrophysical observations of structure~\cite{Feng:2009hw}.

%
\begin{table*}[t]
\begin{center}
\begin{tabular}{lcclc}
\hline
&{\bf Case 1: Asymmetric DM from baryogenesis}&& &{\bf Case 2: Baryogenesis from Asymmetric DM}\\
\hline 
{\bf (1)} &$B-L$ is generated and converted to an $X_2$ asymmetry & \; \; \;  & {\bf (1)}&  $X_1$ charge is generated and produces an $H$ charge\\
& via $(H^{\dagger}X_2)^2/\Lambda_2$ operator.&&  &via $y_H\bar{X}_2 X_1H$ operator.  \\
\hline
{\bf (2)}& $X_2$ charge is transferred to dark matter $X_1$ via $X_2 \to X_1 h$  & \; \; \;  & {\bf (2)} & $H$ chemical potential generates baryon asymmetry $n_B$. \\
&&&& 1st order phase transition is needed to avoid washout of  $n_B$\\
\hline
{\bf (3)} &$X_1\bar{X}_1$ annihilate.
& \; \; \;  & {\bf (3)} & $X_1 \leftrightarrow \bar{X}_1$ oscillations due to  $|H|^2X_1^2/\Lambda_1$ operator\\
&   DM abundance is entirely controlled by $X_1$ asymmetry & \; \; \;  && erase the large primordial $X_1$ asymmetry.\\
&&&&DM abundance is  controlled by symmetric annihilations \\
\hline
{\bf (4)}& $X_1\leftrightarrow \bar{X}_1$ oscillations turn on after $X_1$ freeze-out& \; \; \;  &{\bf (4)}&DM is symmetric \\
\hline
\end{tabular}
\end{center}
\caption{Summary of the two Higgsogenesis scenarios. In both cases, DM annihilations can occur today. } 
\label{tab:summary}
\end{table*}


{\bf Conclusions:} With the discovery of the Higgs, it is important to ask what role this new boson may play cosmologically.  
In electroweak baryogenesis, the Higgs sector provides nonequilibrium dynamics during the EWPT, while in leptogenesis, the Higgs is crucial for CP-violating decays. 
The purpose of this paper was to investigate potential cosmological aspects of a minimal SM-like Higgs boson within a new framework for generating the dark matter and/or baryon densities of the Universe.
Existing baryogenesis scenarios rely on either generating a baryon asymmetry directly or generating a lepton asymmetry that is reprocessed into baryon number.  Similarly, asymmetric dark matter scenarios often rely on dark matter particles carrying baryon or lepton number which gets interconverted between the two sectors. 
Aside from baryon or lepton number, a Higgs number asymmetry is the third kind of asymmetry that can be generated in the SM.  
 In this work, we have explored a new mechanism where a {\it Higgs asymmetry} in the early Universe generates dark or baryonic matter.  
 In our setup, we do not need to introduce new baryon or lepton number violating interactions, relying instead only on interactions between the Higgs and the dark sector.  
We proposed a simple model, with two new fermions $X_{1,2}$ that couple to the Higgs, to illustrate two different cosmological scenarios.  In case I, asymmetric DM can naturally result from a primordial $B\!-\! L$ asymmetry through Higgs charge transfer. Inversely, in case II, a primordial DM asymmetry can generate a Higgs asymmetry, which in turn generates the $B$ asymmetry without additional sources of $B$ violation beyond EW sphalerons, see Table \ref{tab:summary}.
%
%
It is interesting to note that in case II, no $B-L$ is generated, but by having an asymmetry trapped in the spectator $X_2$, we are biasing sphalerons into generating $B+L$. 
Although the Higgs transfer operator remains inaccessible due to its high scale $\Lambda_2 \gtrsim 10^{11}$ GeV, the $X$ sector can be probed experimentally through the EW couplings of $X_2$ or through the dark mediator $\phi$.  
The Higgs boson does possess a coupling to the dark sector, which can lead to invisible Higgs decays.  This is consistent with LHC results.  It is quite interesting that a model as simple as this (just adding two new colorless fermions) remains unconstrained by the LHC but can still have dramatic implications for relating the dark matter and baryogenesis puzzles.
Higgsogenesis shares many dynamical features with electroweak baryogenesis and leptogenesis (e.g., charge equilibration, oscillations, and collisional damping), and quantum kinetic formalisms developed in those contexts~\cite{Konstandin:2004gy} may be of use here as well.

G.~S. thanks T.~Konstandin for discussions at a preliminary stage of this project.
This work is supported by the ERC Starting Grant Cosmo@LHC and DoE Grant \#DE-SC0007859.

\end{document}